\documentclass[11pt,preprint]{aastex}

\usepackage{longtable}

\newcommand{\etal}{{\it et~al.}}

\begin{document}

\title{The Mineralogical Connection Between M- and K-type Asteroids as Indicated by Polarimetry}

\author{Joseph R. Masiero\altaffilmark{1}, Yuna G. Kwon\altaffilmark{1}, Elena Selmi\altaffilmark{2}, Manaswi Kondapally\altaffilmark{3}}

\altaffiltext{1}{Caltech/IPAC, 1200 E California Blvd, MC 100-22, Pasadena, CA 91125, USA, {\it jmasiero@ipac.caltech.edu}}
\altaffiltext{2}{Oxford University}
\altaffiltext{3}{Arizona State University}

\begin{abstract}

  Polarimetry has the capacity to provide a unique probe of the
  surface properties of asteroids.  Trends in polarization behavior as
  a function of wavelength trace asteroid regolith mineral properties
  that are difficult to probe without measurements \textit{in situ} or
  on returned samples. We present recent results from our ongoing
  survey of near-infrared polarimetric properties of asteroids.  Our
  data reveal a mineralogical link between asteroids in the broader M-
  and K- spectral classes.  In particular, M-type objects (16) Psyche,
  (55) Pandora, (135) Hertha, and (216) Kleopatra show the same
  polarimetric-phase behavior as K-type objects (89) Julia, (221) Eos,
  and (233) Asterope from visible through near-infrared light.  The
  near-infrared behavior for these objects is distinct from other
  classes observed to date, and shows a good match to the polarimetric
  properties of M-type asteroid (21) Lutetia from the visible to the
  near-infrared.  The best link for these objects from laboratory
  polarimetric phase curve measurements is to a troilite-rich
  fine-grained regolith.  Our observations indicate that the M- and
  K-type spectral classes are most likely part of a continuum, with
  the observed spectral differences due to heterogeneity from partial
  differentiation, shock darkening of the surface material, or other
  later evolution of the original parent population.  We also provide
  incidental J- and H-band polarimetric observations of other Main
  Belt asteroids obtained during our survey.
\end{abstract}

\section{Introduction}

One of the fundamental outstanding questions in planetary science is:
What is the history of protoplanetary differentiation in the Solar
system, and how many bodies underwent complete melting during
formation?  Observations of asteroids' sizes and masses have revealed
$\sim10$ objects with densities consistent with iron-rich phases of a
differentiated body \citep{carry12}.  Radar observations show
reflectivies for objects like (16) Psyche and (216) Kleopatra are also
consistent with metal-rich surfaces \citep{ostro00,neese20}.  In
contrast with this, isotopic analysis of iron meteorites (another line
of evidence for differentiation in the early Solar system) indicates
that they sample many dozens of sources \citep{burbine02}, from a wide
range of heliocentric distances \citep{zhang24}.

These various observations could all be consistent with a large
original population of objects that differentiated early in the Solar
system's history and then were nearly all shattered before the
majority of the material was removed.  However, there is a notable
absence of the other collisional products that would be expected to
have been produced in such events, including the iron-poor residual
mantle materials.  Other than Vesta and its asteroid family we see no
evidence for the basaltic material that would form during the process
of differentiation, often referred to as the ``missing mantle''
problem \citep{bell89}.

Spectroscopically, metal-rich asteroids have no significant features
at visible or near-infrared wavelengths
\citep{bus02,demeo09,mahlke22}, while photometrically they tend to
have moderate albedos both at visible and near-infrared wavelengths
that fall between the larger C- and S-complexes
\citep{mainzer11tax,masiero14}.  Polarimetric studies, however, offer
a complementary way of studying these objects.  By investigating the
polarimetric-phase curves as a function of wavelength we can probe
compositional information that cannot be revealed from spectra or
colors alone.

As sunlight interacts with asteroid surface materials during
scattering, it can be imparted with a polarization.  This
polarization, which is usually below $2\%$ for Main Belt asteroids,
depends on the phase angle associated with scattering (the
Sun-object-Earth angle), as well as the specific composition and
distribution of the scattering elements on the surface
\citep{muinonen02,masiero09}.  The measurement of the
polarization-phase curve of an asteroid can thus be used to probe the
surface composition of the object in a way that is independent of, but
complementary to, the results provided by photometry and spectroscopy
\citep[see ][for a recent review]{bagnulo24}.
The visible to near-infrared wavelength regime spans the typical
primary grain sizes expected asteroid regolith, and polarization across
these wavelengths can show significant changes tied to the surface
properties \citep{masiero23,bach24}.

In this work, we present recent results from our ongoing survey of
asteroid near-infrared polarimetric phase curves.  We focus on
representative objects from two taxonomic classes: the M- and
K-classes, both of which show muted spectra and moderate albedos.  As
shown below our observations demonstrate a link between these two
classes which stand apart from the behavior of other taxonomies
observed to date.

\section{Observations and Data Reduction}

We obtained new near-infrared polarimetric observations of our targets
using the Wide-field InfraRed Camera+Polarimeter (WIRC+Pol) on the
Palomar 200-inch telescope \citep{tinyanont19a}.  WIRC+Pol uses a
combination quarter-wave plate and polarizing grating to
simultaneously obtain spectral traces of the four linear Stokes
components (+Q, -Q, +U and -U) at both $J$ and $H$ bands ($1.25~\mu$m
and $1.64~\mu$m, respectively).  An up-stream half-wave plate allows
the Stokes vector sampled by each beam to be swapped, allowing for
improved measurement precision, with accuracies of order $0.1\%$
\citep{tinyanont19b}.

Our observing strategy follows the process described in
\citet{masiero22} and \citet{masiero23}.  Following those works, we
generally used A-B pointing positions for each target to enable
improved background subtraction and we obtain a minimum of four
sequences of four half-wave plate rotations per pointing per band to
mitigate any systematics.  The exposure time for each image was scaled
to the predicted brightness of the target at the time of observation,
usually either 1 second (the minimum possible exposure time), 5
seconds, or 30 seconds.  For all observations the telescope was
tracked at non-sidereal rates to maintain the target on approximately
the same pixels for each pointing at all half-wave plate orientations.

We use the standard WIRC+Pol data reduction pipeline
software\footnote{\textit{https://github.com/WIRC-Pol/wirc$\_$drp}} to
correct the images for dark and flat-field effects, and to extract the
degree and angle of polarization for each target.  Dark frames were
obtained at each of the exposure times used at the beginning and end
of each night, and flat field images were obtained in both bands at
the beginning and end of the night as well.  Although WIRC+Pol obtains
polarization measurements at a low spectral resolution across the
entire bandpass, we perform an error-weighted coadd to derive a single
measurement at each band to improve accuracy. Data from all obtained
exposures of an object in a given band on each night are combined to
produce the final polarization measurement of that object.  Based on
previous analyses we add $0.1\%$ in quadrature to the statistical
uncertainty to account for residual systematic errors
\citep{masiero22}, and add $7.8^\circ$ to the derived angle of
polarization in both $J$ and $H$ bands to correct the observed offset
\citep{masiero23}. Our final calibrated polarization measurements for
the M- and K-type objects discussed here are presented in
Table~\ref{tab.astdata}, while incidental observations obtained for
other Main Belt asteroids are given in Table~\ref{tab.appendix}.

\begin{center}
\scriptsize
\noindent
{\tiny
  \begin{longtable}{cccrrcccr}
    \caption{WIRC+Pol M- and K-type Results}\label{tab.astdata}\\
\hline\hline

Asteroid & Observation Date & UT & Total t$_{exp}$ (sec) & Phase (deg) & Filter & $P_r$ & $\theta_r$\\

\hline\hline
\endfirsthead
\caption[]{(continued)}\\
\hline\hline
Asteroid & Observation Date & UT & Total t$_{exp}$ (sec) & Phase (deg) & Filter & $P_r$ & $\theta_r$\\

\hline\hline
\endhead
\hline
\endfoot

16       & 2019-08-30 & 05:22 & 2100 &  9.7 & J & $-1.34 \pm 0.10\%$ & $ 92.2 \pm   2.2$\\ 
16       & 2021-02-03 & 04:23 & 1200 & 18.9 & J & $-0.58 \pm 0.10\%$ & $ 89.4 \pm   5.1$\\ 
16       & 2021-02-03 & 04:32 & 1200 & 18.9 & H & $-0.56 \pm 0.10\%$ & $ 92.7 \pm   5.2$\\ 
16       & 2023-01-07 & 13:30 & 160 & 15.1 & J & $-1.08 \pm 0.11\%$ & $ 89.4 \pm   2.8$\\ 
16       & 2023-01-07 & 13:39 & 160 & 15.1 & H & $-1.18 \pm 0.20\%$ & $ 85.2 \pm   2.8$\\ 
16       & 2023-05-16 & 07:25 & 160 &  2.5 & J & $-0.86 \pm 0.10\%$ & $ 96.2 \pm   3.7$\\ 
16       & 2023-05-16 & 07:34 & 140 &  2.5 & H & $-0.56 \pm 0.10\%$ & $ 83.0 \pm   5.5$\\ 
16       & 2024-05-19 & 11:14 & 160 & 20.6 & J & $-0.35 \pm 0.10\%$ & $ 85.2 \pm   8.6$\\ 
16       & 2024-05-19 & 11:23 & 160 & 20.6 & H & $-0.44 \pm 0.10\%$ & $ 85.0 \pm   6.7$\\ 
16       & 2024-07-22 & 06:00 & 160 &  6.4 & J & $-1.29 \pm 0.10\%$ & $ 92.8 \pm   2.3$\\ 
16       & 2024-07-22 & 06:09 & 160 &  6.4 & H & $-0.93 \pm 0.10\%$ & $ 82.0 \pm   3.2$\\ 
21       & 2025-03-01 & 09:13 & 165 &  6.1 & J & $-1.34 \pm 0.18\%$ & $ 81.7 \pm   3.4$\\ 
21       & 2025-03-01 & 09:21 & 145 &  6.1 & H & $-1.44 \pm 0.17\%$ & $ 86.1 \pm   3.2$\\ 
21       & 2025-03-02 & 09:26 & 160 &  6.5 & J & $-1.49 \pm 0.11\%$ & $ 82.9 \pm   2.1$\\ 
21       & 2025-03-02 & 09:34 & 160 &  6.5 & H & $-1.40 \pm 0.11\%$ & $ 83.1 \pm   2.2$\\ 
21       & 2025-04-12 & 03:29 & 160 & 18.4 & J & $-0.66 \pm 0.12\%$ & $ 75.6 \pm   4.5$\\ 
21       & 2025-04-12 & 03:43 & 160 & 18.4 & H & $-0.78 \pm 0.12\%$ & $ 94.8 \pm   4.0$\\ 
21       & 2025-05-28 & 04:04 & 320 & 20.7 & J & $-0.57 \pm 0.14\%$ & $109.5 \pm   8.6$\\ 
21       & 2025-05-28 & 04:11 & 320 & 20.7 & H & $-0.61 \pm 0.15\%$ & $101.8 \pm   7.1$\\ 
55       & 2025-03-02 & 09:16 & 160 & 13.9 & H & $-1.10 \pm 0.15\%$ & $ 83.9 \pm   2.8$\\ 
89       & 2023-01-14 & 10:01 & 160 &  4.9 & H & $-0.84 \pm 0.10\%$ & $ 95.1 \pm   3.5$\\
135      & 2025-03-02 & 08:29 & 960 & 10.3 & J & $-1.24 \pm 0.11\%$ & $ 81.7 \pm   2.4$\\ 
135      & 2025-03-02 & 08:51 & 960 & 10.3 & H & $-1.45 \pm 0.11\%$ & $ 89.5 \pm   2.0$\\ 
216      & 2023-01-14 & 03:23 & 160 & 26.5 & J & $+0.72 \pm 0.10\%$ & $  1.6 \pm   4.2$\\ 
216      & 2023-01-14 & 03:31 & 160 & 26.5 & H & $+0.64 \pm 0.10\%$ & $176.1 \pm   4.7$\\ 
216      & 2024-01-29 & 08:10 & 160 &  9.4 & J & $-1.16 \pm 0.10\%$ & $ 91.8 \pm   2.8$\\ 
216      & 2024-01-29 & 08:20 & 160 &  9.4 & H & $-1.03 \pm 0.10\%$ & $ 88.6 \pm   3.1$\\ 
216      & 2024-05-19 & 03:57 & 160 & 19.1 & J & $-0.65 \pm 0.24\%$ & $ 97.2 \pm  10.0$\\ 
216      & 2024-05-19 & 04:06 & 160 & 19.1 & H & $-0.66 \pm 0.13\%$ & $120.9 \pm  12.8$\\ 
216      & 2025-05-28 & 04:41 & 160 & 11.4 & J & $-0.85 \pm 0.12\%$ & $ 87.2 \pm   3.7$\\ 
216      & 2025-05-28 & 04:49 & 160 & 11.4 & H & $-1.12 \pm 0.14\%$ & $ 88.3 \pm   2.9$\\ 
221      & 2021-11-08 & 09:18 & 2880 &  7.6 & H & $-1.31 \pm 0.10\%$ & $ 87.0 \pm   2.2$\\ 
221      & 2021-11-08 & 09:29 & 2880 &  7.6 & J & $-1.16 \pm 0.10\%$ & $ 88.5 \pm   2.5$\\ 
221      & 2022-12-01 & 10:22 & 960 & 16.8 & H & $-1.31 \pm 0.11\%$ & $ 92.1 \pm   2.3$\\ 
221      & 2022-12-01 & 10:33 & 960 & 16.8 & J & $-1.05 \pm 0.11\%$ & $ 84.0 \pm   2.8$\\ 
221      & 2023-01-07 & 09:27 & 960 &  9.6 & J & $-1.12 \pm 0.10\%$ & $ 90.5 \pm   2.7$\\ 
221      & 2023-01-07 & 09:50 & 960 &  9.6 & H & $-1.32 \pm 0.11\%$ & $ 90.5 \pm   2.2$\\ 
221      & 2024-01-29 & 13:08 & 960 & 17.4 & J & $-0.89 \pm 0.12\%$ & $ 89.8 \pm   3.5$\\ 
221      & 2024-01-29 & 13:30 & 960 & 17.4 & H & $-0.90 \pm 0.16\%$ & $ 87.7 \pm   3.6$\\ 
221      & 2024-05-19 & 08:49 & 160 & 12.8 & J & $-1.30 \pm 0.14\%$ & $ 89.3 \pm   3.4$\\ 
221      & 2024-05-19 & 08:57 & 160 & 12.8 & H & $-1.28 \pm 0.13\%$ & $ 86.9 \pm   3.2$\\ 
221      & 2025-05-28 & 09:31 & 320 & 16.4 & J & $-1.08 \pm 0.11\%$ & $ 89.5 \pm   2.9$\\ 
221      & 2025-05-28 & 09:39 & 320 & 16.4 & H & $-1.13 \pm 0.11\%$ & $ 88.5 \pm   2.9$\\
233      & 2023-07-26 & 09:40 & 160 & 22.6 & J & $-0.32 \pm 0.11\%$ & $110.3 \pm   9.6$\\ 
233      & 2023-07-26 & 09:49 & 160 & 22.6 & H & $-0.26 \pm 0.10\%$ & $ 73.7 \pm  12.2$\\ 
233      & 2023-08-26 & 07:12 & 160 & 14.4 & J & $-1.48 \pm 0.10\%$ & $ 92.5 \pm   2.1$\\ 
233      & 2023-08-26 & 07:20 & 160 & 14.4 & H & $-1.34 \pm 0.10\%$ & $ 87.4 \pm   2.4$\\ 
233      & 2025-03-02 & 07:42 & 960 & 15.9 & J & $-1.22 \pm 0.11\%$ & $ 86.6 \pm   2.4$\\ 
233      & 2025-03-02 & 08:05 & 960 & 16.0 & H & $-1.20 \pm 0.11\%$ & $ 93.1 \pm   2.4$\\

\hline
  \end{longtable}
The degree of polarization $P_r$ has been rotated so that positive
values represent polarization perpendicular to the
Sun-Asteroid-Telescope scattering plane \textbf{at the time of the
  observations} and negative values represent polarization in this
plane. $\theta_r$ is the \textbf{measured} angle of polarization after
correcting for the observed $7.8^\circ$ systematic offset \textbf{and
  rotated so that $0^\circ$ is aligned with the vector perpendicular
  to the scattering plane}.
}
 \end{center}


\section{Results}

\begin{figure}[ht]
  \centering
\includegraphics[width=0.7\textwidth]{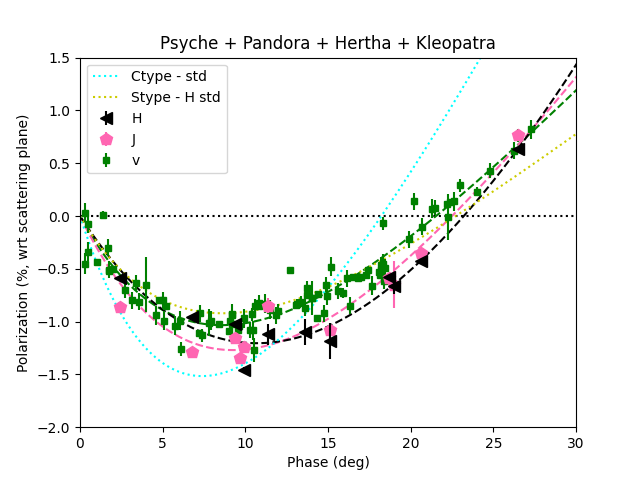}
\caption{Polarimetric phase curves for M-type asteroids (16) Psyche,
  (55) Pandora, (135) Hertha, and (216) Kleopatra.  J- and H-band
  observations were obtained using Palomar/WIRC+Pol while the visible
  bands are from \citet{apd}. Points show the measurements, while the
  lines are best-fit polarization phase curves based on theoretical
  scattering models.  The light-blue and light-yellow lines show the
  best fit curves to the NIR polarization-phase relationships for
  C-types at all wavelengths and S-complex objects a H band
  (respectively, noted as ``Ctype - std'' and ``Stype - H std'') as
  derived by \citet{masiero22} for comparison.  }
\label{fig.mtype}
\end{figure}

We fit our observations with the polarimetric-phase relationship
described in \citet{muinonen09}, following the same implementation
described in \citet{masiero23}. The near-infrared polarimetric phase
curves for M-type asteroids show a change in behavior from the V-band
polarimetric phase curves fitted to data from the literature
\citep{apd}\footnote{Additional V-band observations for Psyche,
Hertha, and Kleopatra were obtained from the Grupo de Ciencias
Planetarias San Juan website:
\textit{http://gcpsj.sdf-eu.org/catalogo.html}}.  Combining our
observations for asteroids (16) Psyche, (55) Pandora, (135) Hertha,
and (216) Kleopatra, we see a slightly deeper negative polarization
branch ($P_{min}$), a higher angle of minimum polarization
($\alpha_{min}$), and a steeper slope at the inversion angle ($h$)
(Figure~\ref{fig.mtype}), although our limited coverage of the
positive polarization branch means that the slope differences may be
due to the limited dataset.  Fit values for the polarimetric phase
curves are given in Table~\ref{tab.fits} with $16^{th}$ and $84^{th}$
percentile uncertainties determined by looking at the range of fits
for 100 Monte Carlo simulations of the polarimetric measurements
assuming the uncertainties follow a Gaussian behavior. Previous
near-infrared polarimetric results \citep{masiero22,masiero23} have
shown that S-type asteroids show subtle changes from visual to NIR and
the ``Barbarian'' asteroids show significant changes, but neither
class of objects shows changes similar to those observed here.

\begin{figure}[ht]
\centering
\includegraphics[width=0.7\textwidth]{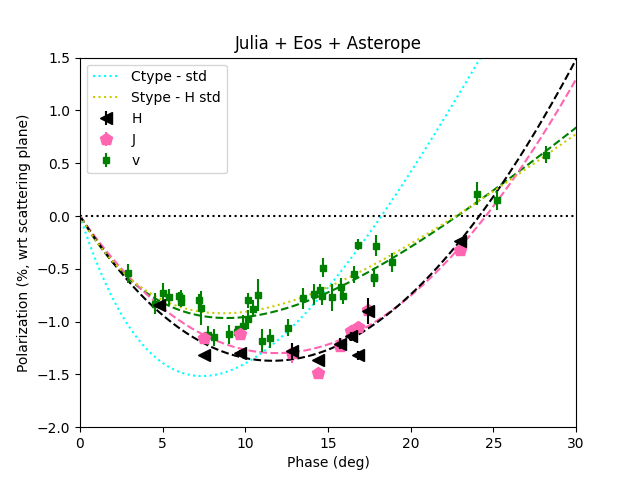}
\caption{The same as Figure~\ref{fig.mtype}, showing polarimetric phase
  curves for K-type asteroids (89) Julia, (221) Eos, and (233) Asterope.}
\label{fig.ktype}
\end{figure}

\begin{figure}[ht]
\centering
\includegraphics[width=0.7\textwidth]{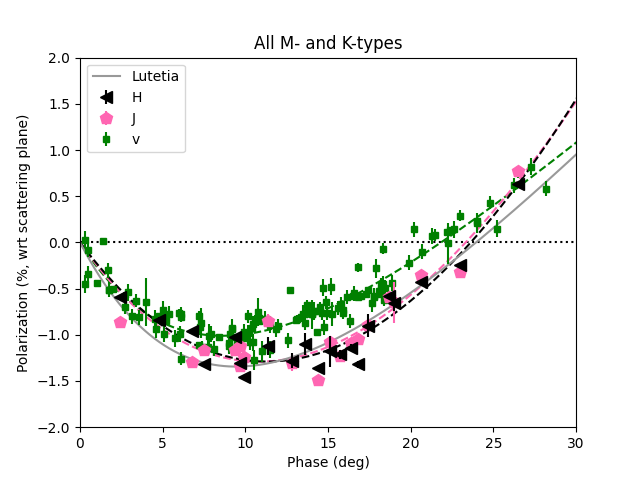}
\caption{The same as Figure~\ref{fig.mtype}, showing the combined data
  for M- and K-types, the best fits to these data, with the best-fit
  curve to the (21) Lutetia V-band data overlaid.}
\label{fig.joint}
\end{figure}

However, we find another asteroid taxonomy that is consistent with the
M-type asteroid polarimetric phase curves at both visible and
near-infrared wavelengths: the K-type asteroids (as shown in
Table~\ref{tab.fits}).  Figure~\ref{fig.ktype} shows the polarimetric
phase curves for asteroids (89) Julia, (221) Eos, and (233) Asterope,
while Fig~\ref{fig.joint} shows the M- and K- type objects plotted
jointly.  (221) Eos and its associated collisional family have
previously been shown to have unique properties at infrared
wavelengths in the form of a $3.4\mu$m albedo that is distinct from
other observed families \citep{masiero14} that has been interpreted to
be a result of shock-darkening of the surface \citep{masiero15}.  This
is consistent with recent results from \citet{sanchez25} showing that
increasing quantities of shock-darkened material added to meteorites
will result in lower albedos across all wavelengths.

\citet{belskaya17} presented fits to the V-band polarimetric phase
curves for both the M- and K- classes. Their fitted $P_{min}$ and
$\alpha_{min}$ are consistent with our fits to the full V-band
literature dataset shown in Table~\ref{tab.fits}, while their
$\alpha_0$ and $h$ slopes are offset from our fits, likely due to
differences in coverage of the positive polarization branch.
Alternatively, as the \citet{belskaya17} data include a larger set of
objects for both classes, the difference in fits may indicate the
presence of compositional (and thus polarimetric) variations within
each class.

\begin{table}[ht]
\begin{center}
  \noindent
  \caption{Polarization Phase Curve Fits}
  \vspace{1ex}
  {
  \noindent
  \begin{tabular}{ccccc}
    \tableline
    $\textbf{M-type}$& & & & \\
  \tableline
Band & $P_{min}$ ($\%$) & $\alpha_{min}$ (deg) & $h (\%/deg)$ & $\alpha_0$ (deg) \\
  \tableline
V & -1.03$^{+0.01}_{-0.01}$ & 8.5$^{+0.2}_{-0.2}$  & 0.13$^{+0.01}_{-0.01}$ & 21.5$^{+0.1}_{-0.1}$ \\
J & -1.27$^{+0.02}_{-0.02}$ & 9.3$^{+0.5}_{-0.5}$  & 0.16$^{+0.01}_{-0.01}$ & 22.5$^{+0.2}_{-0.3}$ \\
H & -1.21$^{+0.04}_{-0.04}$ & 10.7$^{+0.3}_{-0.4}$ & 0.18$^{+0.01}_{-0.01}$ & 23.2$^{+0.1}_{-0.1}$ \\
\hline
  \tableline
    $\textbf{K-type}$& & & & \\
  \tableline
V & -0.97$^{+0.02}_{-0.02}$ & 8.9$^{+0.3}_{-0.3}$  & 0.11$^{+0.01}_{-0.01}$ & 22.8$^{+0.3}_{-0.3}$ \\
J & -1.30$^{+0.02}_{-0.02}$ & 11.9$^{+0.2}_{-0.2}$ & 0.20$^{+0.01}_{-0.01}$ & 24.5$^{+0.4}_{-0.3}$ \\
H & -1.37$^{+0.02}_{-0.02}$ & 11.7$^{+0.2}_{-0.2}$ & 0.21$^{+0.01}_{-0.01}$ & 24.2$^{+0.1}_{-0.1}$ \\
\hline
  \tableline
    $\textbf{M- and K-types fit jointly}$& & & & \\
  \tableline
V & -1.01$^{+0.01}_{-0.01}$ & 8.5$^{+0.2}_{-0.1}$  & 0.12$^{+0.01}_{-0.01}$ & 21.8$^{+0.1}_{-0.1}$ \\
J & -1.29$^{+0.01}_{-0.01}$ & 10.9$^{+0.2}_{-0.3}$ & 0.19$^{+0.01}_{-0.01}$   & 23.3$^{+0.1}_{-0.1}$ \\
H & -1.29$^{+0.02}_{-0.02}$ & 11.3$^{+0.2}_{-0.1}$ & 0.20$^{+0.01}_{-0.01}$   & 23.6$^{+0.1}_{-0.1}$ \\
\hline
  \tableline
    $\textbf{(21) Lutetia}$& & & & \\
  \tableline
V & -1.34$^{+0.02}_{-0.02}$ & 9.5$^{+0.2}_{-0.3}$ & 0.15$^{+0.01}_{-0.01}$ & 23.9$^{+0.2}_{-0.3}$ \\
\hline
  \end{tabular}
  }
  \label{tab.fits}
\end{center}
  Parameters of the best-fit polarization phase curve: $P_{min}$ - the
  negative extrema of the curve; $\alpha_{min}$ - the phase angle
  where the curve reaches $P_{min}$; $h$ - the slope of the curve at
  the inversion angle; $\alpha_0$ - the inversion angle of the curve
  (where the curve crosses $P=0$). $16^{th}$ and $84^{th}$ percentile
  uncertaintes, determined by fits to Monte Carlo simulations of the
  polarimetric measurements, are provided for each parameter.
\end{table}

\section{Discussion}

\begin{figure}[ht]
  \centering
\includegraphics[width=0.7\textwidth]{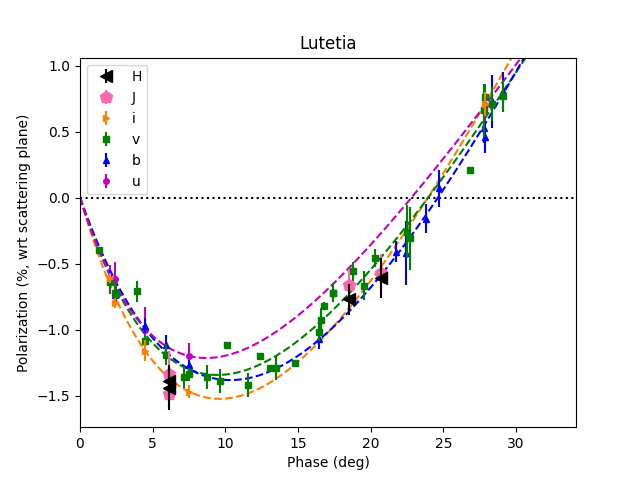}
\caption{The same as Figure~\ref{fig.mtype}, showing polarimetric phase
  curves for M-type asteroid (21) Lutetia.}
\label{fig.lutetia}
\end{figure}

During our analysis we found that our NIR phase curves closely matched
the V-band polarimetric phase curves for the M-type asteroid (21)
Lutetia (fit parameters are shown in Table~\ref{tab.fits} and best-fit
phase curve is shown in Figure~\ref{fig.joint}).  Lutetia was a flyby
target for the ESA mission Rosetta \citep{sierks11} which found a
high-density, low-porosity body with an ancient surface that implies
it has survived intact for the age of the Solar system.  Spectral
measurements of Lutetia showed an absence of the $1-$ and $2-$micron
features that are associated with silicate compositions
\citep{coradini11}.  Lutetia is comparable in size to Eos, Asterope,
and Pandora (sizes of $\sim100~$km).  We obtained NIR polarimetric
observations for Lutetia, which show a similar behavior to the $V$-
and $i$-band data from the literature, as shown
Figure~\ref{fig.lutetia}).  The smaller change seen for Lutetia's
polarimetric phase curve across wavelengths may be due to a difference
in retained grain size distribution of the regolith, as was suggested
to explain the effects seen for the ``Barbarians'' \citep{masiero23}.

\begin{figure}[ht]
  \centering
\includegraphics[width=0.7\textwidth]{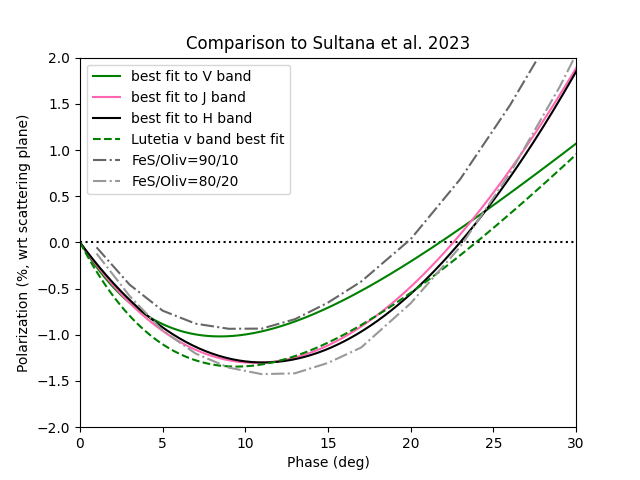}
\caption{The best-fit V-, J-, and H-band curves shown in
  Figure~\ref{fig.joint} compared to laboratory measurements of
  troilite-olivine mixtures from
  \citet{sultana23data}. FeS/Oliv=90/10 is a mixture of $90\%$
  troilite and $10\%$ olivine, while FeS/Oliv=80/20 is a $80\%$/$20\%$
  mixture.  See \citet{sultana23} for more details on the sample materials.}
\label{fig.sult_comp}
\end{figure}

Comparing to laboratory spectra, our J- and H-band measurements are
also consistent with the polarization-phase curve for the $20\%$
olivine-$80\%$ troilite mixture obtained by \citet{sultana23} as shown
in Figure~\ref{fig.sult_comp}, while the V-band curve more closely
matches their $10\%$ olivine-$90\%$ troilite mixture.  This may
indicate that the iron-sulfide troilite (FeS) in the objects we
observed is generally finer-grained than the silicate and so is
contributing less to the near-infrared polarimetric properties, yet
still dominates the surface reflectance.  We would then interpret
Lutetia as having a comparable size distribution for the troilite
and silicate components of its regolith in the size range
  probed by our observations.

It should be noted that the geometric albedo that \citet{sultana23}
determined for these mixtures ($p_V\sim5\%$) is significantly lower
than what is observed for M- and K-type asteroids.  This may be due to
the exclusive use of sub-$\mu$m materials used to build the aggregates
tested, compared with real regoliths that would have a range of
particle sizes.  If the polarization effects are dominated by
different components of the surface than those that produce the albedo
effects, this could explain the discrepency seen here.

The K-type taxonomic class was first identified by \citet{bus02} as
showing muted silicate-like absorption features and a less-red slope.
The canonical member of this class is (221) Eos, the largest object of
a populous, moderate-albedo family in the outer main belt
\citep{masiero13}.  The asteroid (298) Baptistina has also been
associated with the M, K, or Xk class, depending of the taxonomic
scheme \citep{reddy14,mahlke22} based on its moderate albedo and weak
1~and $2~\mu m$ absorption features, which, is a good match for a
mixture of silicate rich LL-chondrites and impact melted material.
Baptistina and Eos also have very similar albedos at $3.4~\mu$m (the
WISE W1 band) of $p_{3.4\mu m}=0.193\pm0.04$ and $p_{3.4\mu
  m}=0.194\pm0.03$, respectively \citep{masiero14,mainzer19}.  NEOWISE
collected these $3.4~\mu$m measurements simultaneously with the
thermal emission measurements used to determine diameter, meaning that
albedos determined in this way are subject to significantly fewer
systematic uncertainties than visual albedos inferred from $H_V$
values.  (The $3.4\mu$m albedos for our other targets discussed here
are also in the moderate region between the bulk of C- and S-complex
objects as fit by \citet{masiero14}: (233) Asterope has $p_{3.4\mu
  m}=0.14\pm0.02$, (75) Eurydike has $p_{3.4\mu m}=0.19\pm0.01$, and
(216) Kleopatra has $p_{3.4\mu m}=0.23\pm0.02$.)

The spectrum of Baptistina has been shown to be a good match to the
Chelyabinsk meteorite \citep{reddy14}. Chelyabinsk shows clear
evidence of having undergone a catastrophic shocking process in its
past, likely a massive collision between $\sim100~$km-scale asteroids
resulting in a complete shattering of the parent body.  This darkening
process involves melting and migration of the troilite in ther mineral
matrix \citep{kohout20}, indicating that the parent body had
experienced some level of heating prior to the impact.  Troilite has a
moderate albedo and featureless spectrum \citep{britt92}, which would
be expected to mute any silicate features of an intermixed surface
material.

ALMA observations of M-type objects (16) Psyche \citep{deKleer21} and
(22) Kalliope \citep{deKleer24} at sub-mm wavelengths show that the
surface materials of these bodies have a moderate emissivity that is
consisitent with metal in mineral phases such as sulfides or oxides.
One complicating factor with these observations is the unexpectedly
low polarization of the sub-mm emission, which these works interpret
as large grains of pure metal ($\>100~\mu$m) in the surface material
that scatter and depolarize the sub-mm emission.  These M-type objects
having a surface dominated by troilite with significant pure metal
inclusions within the surface layer would be consistent with the ALMA
observations.

The polarimetric link we observe here suggests that despite different
spectral taxonomic classifications, the K- and M-type asteroids may in
fact be mineralogically related, and have undergone the same
geophysical processes.  Taxonomy determined from spectral observations
is most strongly affected by the mineralogical absorption features
from iron at $1~\mu$m and $2~\mu$m \citep{busAIII}, while polarimetry
is able to probe different properties of the mineral matrix like
scattering element size, physical grain size, and index of refraction,
which offer a more complete view of the composition of these objects
\citep{belskaya15}.

We note that alternate taxonomic systems place all of the objects
discussed here in similar regions of principal component space, even
if the assigned taxonomic identifier is different. For example,
\citep{demeo09} classify these objects as Xk (Psyche, Asterope), K
(Eos), Xe (Kleopatra) and Xc (Lutetia).  The Tholen classification
system (which included albedo) classifies Psyche, Lutetia, and
Kleopatra as M-type, with Eos as an S-type and Asterope as a T-type.
(298) Baptistina has a Xc taxonomy in the Bus taxonomic system
\citep{lazzaro04}, while \citet{mahlke22} place all our targets in
their larger M-complex.  All these taxonomic systems support an
interpretation of a compositional link and a continuum of spectral
behavior.

\citet{mothediniz05} proposed that the Eos asteroid family, which is
one of the largest collections of K-type objects, is the remnant of a
partially differentiated parent body.  They show a notable similarity
between the spectrum of (221) Eos and the anomolous, differentiated
meteorite Divnoe. Further analysis by \citet{mothediniz08} of 30
family members suggested a connection to the R-chondrites, a class of
olivine-rich, metal-poor meteorites that have compositions that are
consistent with partial differentiation.  (Although it should be noted
that conversely \citet{clark09} associate K-type asteroids with
CO/CV/CK carbonaceous chondrites).

\citet{vernazza11} compared the spectra of Lutetia obtained from
Rosetta to RELAB meteorite samples, finding that the enstatite
chondrites are the closest match over the optical to near-infrared
range.  These results point to a probable formation region for Lutetia
in the terrestrial planet region, the same location proposed as a
probable source of the differentiated (or partially-differentiated)
parents of the iron meteorites \citep{bottke06}.  

Recently, \citet{mcfadden25} performed a re-analysis of the NEOWISE
using the multi-epoch observations to constrain $3.4~\mu$m (W1)
albedos for a larger number of objects than were possible for
\citet{masiero14}.  Using these fits, we can identify the population
of objects with V- and W1- albedos consistent with the objects
discussed here.  These fits show that (21) Lutetia has
$p_{W1}=0.24\pm0.04$, further supporting the polarimetric link
proposed here.  Other objects with moderate V albedos and W1 albedos
include (51) Nemausa, (97) Klotho, (201) Penelope, and (339) Dorothea,
all of which would benefit from future investigation. We have obtained
observations of Klotho at a single phase angle (see
Table~\ref{tab.appendix}) which show a significantly deeper
polarization than would be expected from the other M-type objects
investigated here.  This may be an outlier, or indicate a difference
in surface properties, making this object particularly interesting for
obtaining more observational data.

Taken together, these measurements then may help us settle the
``missing mantle'' problem: instead of basaltic material from a fully
melted object like Vesta, the detritus from the collisions that
exposed the M-type metal-rich asteroids and the iron meteorites may be
the population of K-type objects currently in our catalogs.  This is
consistent with the link between T-type asteroids and troilite
proposed by \citet{britt92met} (two of the three T-type objects from
that study showing the best link have alternate taxonomic
classifications of M/K classification). Our best interpretation of
these objects would be targets that are rich in troilite, with K-type
objects having been closer to the surface of the parent body than the
present-day M-types.  The spectral features of troilite are consistent
with the moderate optical and near-infrared albedos observed for K-
and M-type asteroids.

\section{Implications for \textit{Psyche}}

The NASA \textit{Psyche} mission launched in 2023 and will be
rendezvousing with the asteroid (16) Psyche in 2029 to perform a
detailed analysis of this metal-rich target \citep{psyche}. The
spacecraft carries a spectrophotometer, a gamma-ray and neutron
spectrometer, and a magnetometer which it will use to map the
topography and measure the internal gravity and surface composition of
the asteroid.  Based on the link that we present here between the K-
and M-type objects, we would anticipate that the results from
\textit{Psyche} will generally show a surface that is a mixture of
shocked silicate materials in a surface dominated by troilite or other
iron-rich minerals, rather than a pure metallic surface.  This would
be consistent with the observations presented by \citet{sanchez17} and
\citet{cantillo21} showing that spectral observations of (16) Psyche
indicate a surface that is composed of primarily an iron-rich material
with $\sim10-15\%$ of pyroxene and chondritic materials, as well as
the ALMA results shown by \citet{deKleer21}.

Troilite is commonly found intermixed in iron meteorites, though it is
signficantly weaker than iron-nickel as so would be expected to
dominate the regolith of a body that has undergone significant
gardening from background impacts over gigayear timescales.  It has a
higher density that the silicates commonly found in meteorites, and so
would contribute to the overall high bulk density of (16) Psyche. As
troilite is non-magnetic it would not be expected to contribute to any
surface magnetic fields that could be detected by the \textit{Psyche}
mission.

\section{Conclusions}

We present new near-infrared polarimetric observations of M- and
K-type asteroids.  Our observations show changes from visible to NIR
wavelengths that are consistent between these two classes, which
indicate a mineralogical link between these populations. Further
observations of more members of these populations, and more complete
near-infrared phase curves for the object shown here will help improve
this link. The associations of these two populations with extreme
collisions in the early Solar system connect the evolution of these
bodies as well as their composition.  Taken together, these features
paint a compelling story for small body evolution, with iron
meteorites and M-type asteroids representing the interiors of
partially differentiated, troilite-rich, collisionally-disrupted
protoplanets, and K-type bodies tracing the material that originated
closer to the surface that was dispersed during these ancient
collisions.

\section*{Acknowledgments}

We thank the referees for their helpful comments that improved this
paper.  JRM thanks Dan Britt, Olivier Poch, and Michele Bannister for
very helpful discussions with interpreting these results, and the
staff of Palomar Observatory for all their assistance over the years
collecting these obsevations.  Based on observations obtained at the
Hale Telescope, Palomar Observatory as part of a continuing
collaboration between the California Institute of Technology,
NASA/JPL, Yale University, and the National Astronomical Observatories
of China.  This research made use of Photutils, an Astropy package for
detection and photometry of astronomical sources (Bradley et al.
2019).  This work made use of
Astropy:\footnote{http://www.astropy.org} a community-developed core
Python package and an ecosystem of tools and resources for astronomy
\citep{astropy2013, astropy2018, astropy2022}.  This publication makes
use of data products from the Wide-field Infrared Survey Explorer,
which is a joint project of the University of California, Los Angeles,
and the Jet Propulsion Laboratory/California Institute of Technology,
funded by the National Aeronautics and Space Administration. This
publication also makes use of data products from NEOWISE, which is a
joint project of University of California, Los Angeles and the Jet
Propulsion Laboratory/California Institute of Technology, funded by
the Planetary Science Division of the National Aeronautics and Space
Administration.  This research has made use of data and services
provided by the International Astronomical Union’s Minor Planet
Center.  This research has made use of NASA’s Astrophysics Data System

\newpage
\appendix
\section*{Appendix A: Incidental Observations}

\begin{figure}[ht]
  \centering
\renewcommand\thefigure{A.1}
\includegraphics[width=0.5\textwidth]{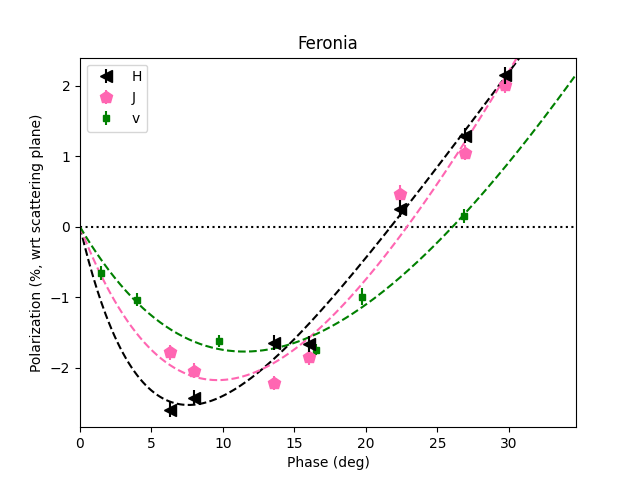}
\caption{The same as Figure~\ref{fig.mtype}, showing polarimetric phase
  curves for T-type asteroid (72) Feronia.}
\label{fig.feronia}
\end{figure}

\begin{figure}[ht]
  \centering
\renewcommand\thefigure{A.2}
\includegraphics[width=0.5\textwidth]{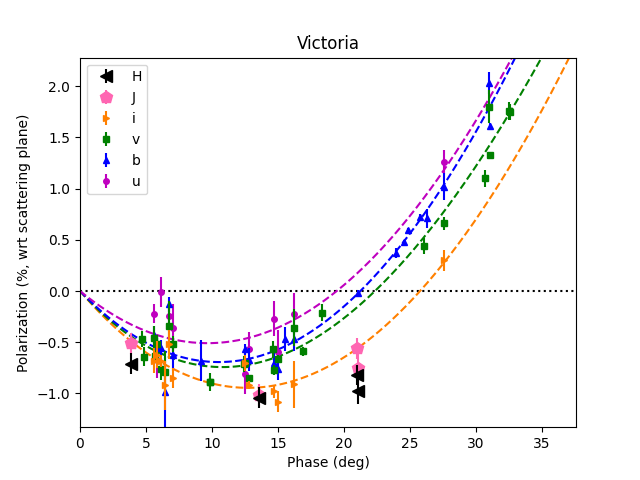}
\caption{The same as Figure~\ref{fig.mtype}, showing polarimetric phase
  curves for S/L-type asteroid (12) Victoria.}
\label{fig.victoria}
\end{figure}

Table~\ref{tab.appendix} presents incidental observations of other
Main Belt asteroids obtained with WIRC+Pol over the course of our
survey.  Observations originally presented in \citet{masiero22} are
revised here with corrected $\theta$ angles of polarization and
polarization uncertainties.  Observations presented in
\citet{masiero23} are unchanged and so not included in this table.
Two notable objects from our sample are shown in
Figs~\ref{fig.feronia} and \ref{fig.victoria}: T-type object (72)
Feronia that shows the deepest P$_{min}$ of any object observed, and
S/L-type object (12) Victoria which shows a continuation to the
near-infrared of the shift in polarization-phase curve observed by
\citet{belskaya09}.

\begin{center}
\renewcommand\thetable{A.1}
\scriptsize
\noindent
{\tiny
\begin{longtable}{cccrrcccr}
\caption{WIRC+Pol Incidental Main Belt Observations}\label{tab.appendix}\\
\hline\hline

Asteroid & Observation Date & UT & Total t$_{exp}$ (sec) & Phase (deg) & Filter & $P_r$ & $\theta_r$\\

\hline\hline
\endfirsthead
\caption[]{(continued)}\\
\hline\hline
Asteroid & Observation Date & UT & Total t$_{exp}$ (sec) & Phase (deg) & Filter & $P_r$ & $\theta_r$\\

\hline\hline
\endhead
\hline
\endfoot

1        & 2021-09-04 & 10:57 & 320 & 21.0 & J & $+0.89 \pm 0.10\%$ & $179.9 \pm   3.3$\\ 
1        & 2021-09-04 & 11:02 & 320 & 21.0 & H & $+0.88 \pm 0.10\%$ & $  7.7 \pm   3.3$\\ 
1        & 2021-11-08 & 10:39 & 32 &  8.2 & H & $-1.59 \pm 0.10\%$ & $ 94.4 \pm   1.8$\\ 
1        & 2021-11-08 & 10:44 & 32 &  8.2 & J & $-1.78 \pm 0.10\%$ & $ 89.4 \pm   1.7$\\ 
1        & 2022-04-02 & 04:51 & 160 & 19.3 & J & $+0.52 \pm 0.10\%$ & $  3.6 \pm   5.7$\\ 
1        & 2023-01-07 & 10:26 & 32 & 22.1 & J & $+1.35 \pm 0.10\%$ & $179.9 \pm   2.2$\\ 
1        & 2023-01-07 & 10:33 & 32 & 22.1 & H & $+1.26 \pm 0.10\%$ & $  0.2 \pm   2.3$\\ 
1        & 2023-05-16 & 05:39 & 32 & 20.1 & J & $+0.38 \pm 0.10\%$ & $  7.3 \pm   7.8$\\ 
1        & 2023-05-16 & 05:46 & 32 & 20.1 & H & $+0.87 \pm 0.10\%$ & $179.6 \pm   3.4$\\ 
1        & 2023-07-26 & 04:01 & 160 & 20.9 & J & $+0.81 \pm 0.10\%$ & $179.7 \pm   3.6$\\ 
1        & 2023-07-26 & 04:10 & 160 & 20.9 & H & $+0.93 \pm 0.10\%$ & $  2.8 \pm   3.1$\\
1        & 2024-05-19 & 11:33 & 32 & 16.3 & J & $-0.51 \pm 0.10\%$ & $ 87.6 \pm   5.7$\\ 
1        & 2024-05-19 & 11:39 & 32 & 16.3 & H & $-0.35 \pm 0.10\%$ & $ 76.7 \pm   8.5$\\ 
2        & 2019-03-17 & 09:32 & 505 & 13.9 & J & $-1.01 \pm 0.10\%$ & $ 94.8 \pm   2.9$\\ 
2        & 2019-03-17 & 10:08 & 510 & 13.9 & H & $-0.80 \pm 0.10\%$ & $ 98.3 \pm   3.7$\\ 
2        & 2021-05-30 & 11:19 & 480 & 17.4 & H & $-0.24 \pm 0.10\%$ & $ 99.1 \pm  12.1$\\ 
2        & 2021-05-30 & 11:24 & 480 & 17.4 & J & $-0.30 \pm 0.10\%$ & $ 87.5 \pm   9.8$\\ 
2        & 2021-06-26 & 10:44 & 32 & 18.0 & H & $-0.17 \pm 0.10\%$ & $ 65.7 \pm  18.3$\\ 
2        & 2021-06-26 & 10:48 & 32 & 18.0 & J & $-0.16 \pm 0.10\%$ & $ 67.5 \pm  20.2$\\ 
2        & 2021-09-04 & 07:32 & 160 &  3.2 & H & $-1.35 \pm 0.10\%$ & $ 90.0 \pm   2.2$\\ 
2        & 2021-09-04 & 07:37 & 160 &  3.2 & J & $-1.27 \pm 0.10\%$ & $ 91.9 \pm   2.3$\\ 
2        & 2021-11-08 & 04:21 & 160 & 17.4 & H & $-0.26 \pm 0.10\%$ & $ 80.2 \pm  11.6$\\ 
2        & 2021-11-08 & 04:26 & 160 & 17.4 & J & $-0.11 \pm 0.10\%$ & $ 87.8 \pm  25.8$\\ 
2        & 2022-10-15 & 12:10 & 24 & 26.0 & H & $+2.00 \pm 0.10\%$ & $  1.3 \pm   1.5$\\ 
2        & 2024-01-29 & 12:19 & 160 & 20.5 & J & $+0.50 \pm 0.10\%$ & $  4.0 \pm   5.9$\\ 
2        & 2024-01-29 & 12:28 & 160 & 20.5 & H & $+0.62 \pm 0.10\%$ & $  2.5 \pm   4.7$\\ 
2        & 2024-05-19 & 09:45 & 160 & 14.5 & J & $-0.72 \pm 0.10\%$ & $ 87.0 \pm   4.1$\\ 
2        & 2024-05-19 & 09:54 & 160 & 14.5 & H & $-0.86 \pm 0.10\%$ & $ 88.8 \pm   3.4$\\ 
2        & 2025-04-12 & 11:54 & 160 & 16.0 & J & $-0.73 \pm 0.10\%$ & $ 85.7 \pm   4.1$\\ 
2        & 2025-04-12 & 12:03 & 160 & 16.0 & H & $-0.62 \pm 0.10\%$ & $ 88.8 \pm   4.7$\\ 
2        & 2025-05-28 & 09:02 & 160 & 17.0 & J & $-0.42 \pm 0.10\%$ & $ 88.1 \pm   7.1$\\ 
2        & 2025-05-28 & 09:10 & 160 & 17.0 & H & $-0.49 \pm 0.10\%$ & $ 80.6 \pm   6.0$\\ 
3        & 2021-05-30 & 07:53 & 640 &  6.2 & J & $-0.72 \pm 0.10\%$ & $ 88.9 \pm   4.1$\\ 
3        & 2021-05-30 & 07:57 & 645 &  6.2 & H & $-0.95 \pm 0.10\%$ & $ 89.6 \pm   3.1$\\ 
3        & 2021-06-26 & 07:16 & 240 &  8.6 & H & $-0.75 \pm 0.10\%$ & $ 87.4 \pm   3.9$\\ 
3        & 2021-06-26 & 07:21 & 240 &  8.6 & J & $-0.86 \pm 0.10\%$ & $ 95.6 \pm   3.4$\\ 
3        & 2021-09-04 & 03:56 & 120 & 18.4 & H & $-0.45 \pm 0.11\%$ & $100.8 \pm   6.5$\\ 
3        & 2021-09-04 & 03:59 & 120 & 18.4 & J & $-0.47 \pm 0.11\%$ & $ 83.6 \pm   6.3$\\ 
3        & 2022-05-31 & 11:28 & 160 & 23.0 & H & $+0.17 \pm 0.10\%$ & $164.7 \pm  17.4$\\ 
3        & 2022-05-31 & 11:32 & 160 & 23.0 & J & $+0.26 \pm 0.10\%$ & $  9.3 \pm  11.5$\\ 
3        & 2022-07-11 & 09:31 & 160 & 21.2 & H & $-0.16 \pm 0.10\%$ & $ 76.5 \pm  18.5$\\ 
3        & 2022-07-11 & 09:35 & 160 & 21.2 & J & $+0.03 \pm 0.10\%$ & $172.9 \pm  86.3$\\ 
3        & 2022-10-15 & 04:57 & 160 & 18.1 & H & $-0.41 \pm 0.10\%$ & $ 89.1 \pm   7.0$\\ 
3        & 2022-10-15 & 05:02 & 160 & 18.1 & J & $-0.33 \pm 0.10\%$ & $ 93.2 \pm   9.0$\\ 
3        & 2023-01-07 & 03:28 & 160 & 26.9 & J & $+0.70 \pm 0.10\%$ & $  4.8 \pm   4.2$\\ 
3        & 2023-01-07 & 03:38 & 160 & 26.9 & H & $+0.39 \pm 0.10\%$ & $179.8 \pm   7.5$\\ 
3        & 2024-01-29 & 09:19 & 160 & 14.8 & J & $-0.44 \pm 0.10\%$ & $ 85.5 \pm   6.7$\\ 
3        & 2024-01-29 & 09:28 & 160 & 14.8 & H & $-0.70 \pm 0.10\%$ & $ 83.0 \pm   4.2$\\ 
4        & 2019-08-30 & 10:13 & 2380 & 23.3 & J & $-0.09 \pm 0.10\%$ & $ 55.6 \pm  32.9$\\ 
4        & 2021-02-03 & 11:38 & 40 & 14.3 & H & $-0.65 \pm 0.11\%$ & $ 89.2 \pm   4.5$\\ 
4        & 2021-02-03 & 11:43 & 41 & 14.3 & J & $-0.31 \pm 0.11\%$ & $ 93.6 \pm   9.6$\\ 
4        & 2021-05-30 & 06:22 & 64 & 26.6 & J & $+0.15 \pm 0.10\%$ & $ 11.0 \pm  19.9$\\ 
4        & 2021-05-30 & 06:26 & 64 & 26.6 & H & $+0.25 \pm 0.10\%$ & $  3.6 \pm  11.8$\\ 
4        & 2021-06-26 & 04:15 & 48 & 26.2 & H & $+0.13 \pm 0.10\%$ & $166.8 \pm  22.5$\\ 
4        & 2021-06-26 & 04:19 & 48 & 26.2 & J & $+0.18 \pm 0.10\%$ & $ 17.1 \pm  16.0$\\ 
4        & 2022-07-11 & 10:00 & 64 & 19.6 & J & $-0.22 \pm 0.11\%$ & $120.8 \pm  13.3$\\ 
4        & 2022-07-11 & 10:03 & 64 & 19.6 & H & $-0.36 \pm 0.10\%$ & $ 73.6 \pm   8.3$\\ 
4        & 2022-10-15 & 04:19 & 32 & 21.4 & H & $-0.23 \pm 0.10\%$ & $100.5 \pm  12.9$\\ 
4        & 2022-10-15 & 04:23 & 32 & 21.4 & J & $-0.12 \pm 0.10\%$ & $ 97.3 \pm  24.2$\\ 
4        & 2022-12-01 & 03:01 & 32 & 24.3 & H & $-0.03 \pm 0.10\%$ & $ 93.9 \pm 109.0$\\ 
4        & 2022-12-01 & 03:04 & 32 & 24.3 & J & $+0.08 \pm 0.10\%$ & $ 10.0 \pm  35.7$\\ 
4        & 2023-01-14 & 02:20 & 32 & 19.6 & J & $-0.07 \pm 0.10\%$ & $ 67.6 \pm  45.8$\\ 
4        & 2023-01-14 & 02:27 & 32 & 19.6 & H & $-0.20 \pm 0.10\%$ & $ 89.5 \pm  14.5$\\ 
6        & 2023-01-14 & 09:02 & 160 &  6.7 & J & $-0.79 \pm 0.11\%$ & $ 94.5 \pm   4.1$\\ 
6        & 2023-01-14 & 09:11 & 160 &  6.7 & H & $-0.66 \pm 0.12\%$ & $ 85.9 \pm   5.6$\\ 
6        & 2024-01-29 & 11:57 & 160 & 19.7 & J & $-0.34 \pm 0.11\%$ & $ 83.9 \pm   8.6$\\ 
6        & 2024-01-29 & 12:07 & 160 & 19.7 & H & $-0.57 \pm 0.10\%$ & $ 92.2 \pm   5.1$\\ 
7        & 2021-09-04 & 11:45 & 480 & 30.5 & H & $+0.71 \pm 0.10\%$ & $178.4 \pm   4.2$\\ 
7        & 2021-09-04 & 11:49 & 480 & 30.5 & J & $+0.81 \pm 0.10\%$ & $179.7 \pm   3.7$\\ 
7        & 2021-11-08 & 12:32 & 160 & 29.1 & H & $+0.62 \pm 0.10\%$ & $  1.2 \pm   4.7$\\ 
7        & 2021-11-08 & 12:37 & 160 & 29.1 & J & $+0.72 \pm 0.11\%$ & $ 13.1 \pm   4.1$\\ 
7        & 2022-04-02 & 05:48 & 160 & 25.5 & H & $+0.22 \pm 0.10\%$ & $ 10.9 \pm  13.5$\\ 
7        & 2022-04-02 & 05:53 & 160 & 25.5 & J & $+0.18 \pm 0.10\%$ & $170.2 \pm  16.8$\\ 
7        & 2023-01-07 & 13:10 & 160 & 18.7 & J & $-0.39 \pm 0.10\%$ & $ 78.9 \pm   7.6$\\ 
7        & 2023-01-07 & 13:19 & 160 & 18.7 & H & $-0.58 \pm 0.10\%$ & $ 89.0 \pm   5.0$\\ 
7        & 2023-05-16 & 07:05 & 160 &  6.6 & J & $-1.26 \pm 0.12\%$ & $ 88.2 \pm   2.5$\\ 
7        & 2023-05-16 & 07:14 & 160 &  6.6 & H & $-0.96 \pm 0.14\%$ & $ 94.0 \pm   3.1$\\ 
10       & 2022-04-02 & 09:12 & 160 & 10.6 & H & $-1.19 \pm 0.10\%$ & $ 88.1 \pm   2.5$\\ 
10       & 2022-04-02 & 09:16 & 160 & 10.6 & J & $-0.99 \pm 0.10\%$ & $ 88.9 \pm   3.0$\\ 
10       & 2022-05-31 & 04:22 & 160 & 12.4 & H & $-0.92 \pm 0.10\%$ & $ 91.7 \pm   3.2$\\ 
10       & 2022-05-31 & 04:27 & 160 & 12.4 & J & $-0.89 \pm 0.10\%$ & $ 93.5 \pm   3.3$\\ 
10       & 2022-07-11 & 04:22 & 160 & 20.5 & H & $+0.82 \pm 0.10\%$ & $  5.4 \pm   3.6$\\ 
10       & 2022-07-11 & 04:26 & 160 & 20.5 & J & $+0.83 \pm 0.11\%$ & $  0.4 \pm   3.6$\\ 
10       & 2023-07-26 & 07:49 & 160 &  5.8 & J & $-1.45 \pm 0.10\%$ & $ 94.1 \pm   2.1$\\ 
10       & 2023-07-26 & 07:57 & 160 &  5.8 & H & $-0.98 \pm 0.10\%$ & $ 89.0 \pm   3.0$\\ 
10       & 2023-08-26 & 05:25 & 160 &  6.0 & J & $-1.43 \pm 0.10\%$ & $ 87.0 \pm   2.1$\\ 
10       & 2023-08-26 & 05:33 & 160 &  6.0 & H & $-1.08 \pm 0.10\%$ & $ 93.1 \pm   2.7$\\ 
10       & 2024-10-07 & 06:12 & 160 &  5.3 & J & $-1.30 \pm 0.10\%$ & $ 86.6 \pm   2.3$\\ 
10       & 2024-10-07 & 06:20 & 160 &  5.2 & H & $-1.23 \pm 0.10\%$ & $ 84.7 \pm   2.4$\\ 
12       & 2022-12-01 & 12:53 & 160 & 13.6 & H & $-1.05 \pm 0.11\%$ & $ 88.0 \pm   2.8$\\ 
12       & 2022-12-01 & 12:58 & 160 & 13.5 & J & $-1.02 \pm 0.11\%$ & $ 88.4 \pm   2.9$\\ 
12       & 2023-01-07 & 08:13 & 160 &  3.9 & J & $-0.50 \pm 0.10\%$ & $ 90.6 \pm   5.9$\\ 
12       & 2023-01-07 & 08:22 & 160 &  3.9 & H & $-0.71 \pm 0.10\%$ & $ 86.1 \pm   4.2$\\ 
12       & 2024-01-29 & 10:49 & 160 & 21.1 & J & $-0.75 \pm 0.18\%$ & $ 79.2 \pm   4.6$\\ 
12       & 2024-01-29 & 10:59 & 160 & 21.1 & H & $-0.97 \pm 0.12\%$ & $ 84.2 \pm   3.2$\\ 
12       & 2024-05-19 & 06:54 & 160 & 21.0 & J & $-0.55 \pm 0.10\%$ & $ 89.8 \pm   5.3$\\ 
12       & 2024-05-19 & 07:02 & 160 & 21.0 & H & $-0.81 \pm 0.10\%$ & $ 89.7 \pm   3.6$\\ 
15       & 2019-08-30 & 07:46 & 1400 &  9.5 & J & $-0.99 \pm 0.10\%$ & $ 91.6 \pm   3.0$\\ 
15       & 2021-02-03 & 05:47 & 400 &  6.1 & J & $-0.91 \pm 0.10\%$ & $ 91.3 \pm   3.2$\\ 
15       & 2021-02-03 & 05:53 & 400 &  6.1 & H & $-0.80 \pm 0.10\%$ & $ 94.9 \pm   3.6$\\ 
20       & 2021-11-08 & 12:01 & 320 & 28.7 & J & $+0.72 \pm 0.10\%$ & $175.8 \pm   4.1$\\ 
20       & 2021-11-08 & 12:06 & 320 & 28.7 & H & $+0.77 \pm 0.10\%$ & $179.8 \pm   3.8$\\ 
20       & 2022-04-02 & 07:00 & 160 & 23.9 & H & $+0.25 \pm 0.10\%$ & $  2.8 \pm  11.5$\\ 
20       & 2022-04-02 & 07:05 & 160 & 23.9 & J & $+0.22 \pm 0.10\%$ & $167.7 \pm  13.1$\\ 
20       & 2023-07-26 & 06:10 & 160 & 15.2 & J & $-0.41 \pm 0.10\%$ & $ 91.3 \pm   7.2$\\ 
20       & 2023-07-26 & 06:19 & 160 & 15.2 & H & $-0.59 \pm 0.10\%$ & $ 87.9 \pm   4.9$\\ 
20       & 2023-08-26 & 03:23 & 160 & 20.7 & J & $-0.13 \pm 0.11\%$ & $ 72.4 \pm  22.6$\\ 
20       & 2023-08-26 & 03:33 & 160 & 20.7 & H & $-0.24 \pm 0.10\%$ & $118.4 \pm  12.7$\\ 
20       & 2024-10-07 & 06:33 & 320 &  3.9 & H & $-0.64 \pm 0.10\%$ & $ 86.2 \pm   4.6$\\ 
25       & 2021-11-08 & 06:04 & 160 & 11.0 & H & $-0.48 \pm 0.10\%$ & $ 78.9 \pm   6.1$\\ 
25       & 2021-11-08 & 06:08 & 160 & 11.0 & J & $-0.69 \pm 0.10\%$ & $ 87.3 \pm   4.3$\\ 
27       & 2023-01-14 & 04:30 & 160 & 27.6 & J & $+0.43 \pm 0.10\%$ & $ 16.4 \pm   6.8$\\ 
27       & 2023-01-14 & 04:38 & 160 & 27.6 & H & $+0.34 \pm 0.10\%$ & $ 11.5 \pm   8.6$\\ 
27       & 2024-01-29 & 12:39 & 160 & 23.7 & J & $+0.13 \pm 0.10\%$ & $ 28.6 \pm  23.6$\\ 
27       & 2024-01-29 & 12:49 & 160 & 23.7 & H & $+0.33 \pm 0.11\%$ & $  3.9 \pm   9.2$\\ 
30       & 2023-01-14 & 04:50 & 160 & 21.4 & J & $-0.17 \pm 0.10\%$ & $ 74.4 \pm  17.5$\\ 
30       & 2023-01-14 & 04:59 & 160 & 21.4 & H & $-0.31 \pm 0.10\%$ & $ 85.6 \pm   9.3$\\ 
31       & 2022-10-15 & 05:35 & 160 &  2.7 & H & $-0.58 \pm 0.10\%$ & $101.4 \pm   5.0$\\ 
31       & 2022-10-15 & 05:40 & 160 &  2.7 & J & $-0.75 \pm 0.10\%$ & $ 93.2 \pm   4.0$\\ 
31       & 2022-12-01 & 05:59 & 160 & 17.4 & H & $-0.35 \pm 0.10\%$ & $ 84.9 \pm   8.9$\\ 
31       & 2022-12-01 & 06:04 & 160 & 17.4 & J & $-0.57 \pm 0.10\%$ & $ 81.8 \pm   5.3$\\ 
31       & 2023-01-07 & 04:34 & 160 & 21.6 & H & $+0.39 \pm 0.10\%$ & $179.6 \pm   8.1$\\ 
31       & 2023-01-07 & 04:43 & 160 & 21.6 & J & $+0.60 \pm 0.10\%$ & $179.0 \pm   5.3$\\ 
31       & 2024-01-29 & 10:07 & 160 & 14.8 & J & $-1.05 \pm 0.11\%$ & $ 87.8 \pm   3.0$\\ 
31       & 2024-01-29 & 10:16 & 160 & 14.8 & H & $-0.71 \pm 0.11\%$ & $ 89.0 \pm   4.6$\\ 
31       & 2024-05-19 & 07:12 & 160 & 20.8 & H & $+0.24 \pm 0.12\%$ & $  3.2 \pm  13.3$\\ 
31       & 2024-05-19 & 07:20 & 160 & 20.8 & J & $+0.47 \pm 0.11\%$ & $  4.6 \pm   6.7$\\
41       & 2022-04-02 & 11:57 & 160 & 27.5 & H & $+2.49 \pm 0.10\%$ & $  4.5 \pm   1.2$\\ 
41       & 2022-04-02 & 12:01 & 160 & 27.5 & J & $+2.85 \pm 0.10\%$ & $  1.3 \pm   1.0$\\ 
41       & 2022-05-31 & 09:49 & 160 & 14.3 & H & $-1.66 \pm 0.10\%$ & $ 92.9 \pm   1.8$\\ 
41       & 2022-05-31 & 09:54 & 160 & 14.3 & J & $-1.46 \pm 0.10\%$ & $ 93.4 \pm   2.0$\\ 
41       & 2022-07-11 & 07:22 & 160 & 19.3 & H & $-0.07 \pm 0.10\%$ & $111.8 \pm  40.7$\\ 
41       & 2022-07-11 & 07:27 & 160 & 19.3 & J & $-0.17 \pm 0.10\%$ & $ 79.3 \pm  17.5$\\ 
41       & 2023-07-26 & 10:07 & 960 & 17.9 & J & $-0.82 \pm 0.10\%$ & $ 91.4 \pm   3.6$\\ 
41       & 2023-07-26 & 10:30 & 960 & 17.9 & H & $-0.56 \pm 0.10\%$ & $ 76.0 \pm   5.4$\\ 
41       & 2023-08-26 & 07:38 & 960 & 12.2 & J & $-1.81 \pm 0.10\%$ & $ 89.8 \pm   1.6$\\ 
41       & 2023-08-26 & 08:00 & 960 & 12.2 & H & $-1.40 \pm 0.10\%$ & $ 86.8 \pm   2.1$\\ 
44       & 2021-11-08 & 11:28 & 160 & 16.7 & H & $-0.18 \pm 0.10\%$ & $ 53.2 \pm  16.5$\\ 
44       & 2021-11-08 & 11:33 & 160 & 16.7 & J & $-0.22 \pm 0.10\%$ & $ 69.9 \pm  13.7$\\ 
44       & 2022-04-02 & 05:29 & 160 & 28.2 & H & $+0.35 \pm 0.10\%$ & $ 11.3 \pm   8.4$\\ 
44       & 2022-04-02 & 05:33 & 160 & 28.2 & J & $+0.50 \pm 0.10\%$ & $176.1 \pm   5.9$\\ 
44       & 2023-01-07 & 13:49 & 160 & 20.6 & J & $+0.38 \pm 0.12\%$ & $ 39.6 \pm  12.5$\\ 
44       & 2023-01-07 & 13:58 & 160 & 20.6 & H & $+0.05 \pm 0.10\%$ & $ 20.2 \pm  67.6$\\ 
44       & 2023-05-16 & 07:55 & 160 &  3.6 & H & $-0.44 \pm 0.10\%$ & $ 74.7 \pm   7.0$\\ 
46       & 2022-10-15 & 05:55 & 160 &  6.2 & H & $-1.45 \pm 0.10\%$ & $ 90.9 \pm   2.0$\\ 
46       & 2022-10-15 & 05:59 & 160 &  6.2 & J & $-1.82 \pm 0.10\%$ & $ 89.0 \pm   1.6$\\ 
46       & 2022-12-01 & 06:12 & 80 & 17.6 & J & $-0.59 \pm 0.11\%$ & $102.3 \pm   5.1$\\ 
46       & 2023-01-07 & 05:00 & 960 & 25.1 & J & $+2.69 \pm 0.10\%$ & $  1.5 \pm   1.1$\\ 
46       & 2023-01-07 & 05:23 & 960 & 25.1 & H & $+2.05 \pm 0.10\%$ & $  0.7 \pm   1.5$\\ 
46       & 2024-01-29 & 08:36 & 960 &  9.1 & J & $-1.17 \pm 0.11\%$ & $102.1 \pm   2.9$\\ 
46       & 2024-01-29 & 09:00 & 960 &  9.1 & H & $-1.80 \pm 0.14\%$ & $ 88.0 \pm   1.9$\\ 
51       & 2022-05-31 & 07:50 & 160 & 17.6 & H & $-0.55 \pm 0.11\%$ & $ 75.4 \pm   5.3$\\ 
51       & 2022-05-31 & 07:54 & 160 & 17.6 & J & $-0.83 \pm 0.10\%$ & $ 81.6 \pm   3.6$\\ 
51       & 2022-07-11 & 06:14 & 160 & 25.7 & H & $+1.67 \pm 0.11\%$ & $179.9 \pm   1.8$\\ 
51       & 2022-07-11 & 06:18 & 160 & 25.7 & J & $+1.51 \pm 0.11\%$ & $179.9 \pm   2.0$\\ 
51       & 2023-07-26 & 09:02 & 160 & 19.2 & J & $-0.47 \pm 0.11\%$ & $ 93.1 \pm   6.8$\\ 
51       & 2023-07-26 & 09:11 & 160 & 19.2 & H & $-0.09 \pm 0.10\%$ & $ 73.5 \pm  34.0$\\ 
51       & 2023-08-26 & 06:34 & 160 &  8.4 & J & $-1.99 \pm 0.10\%$ & $ 92.2 \pm   1.5$\\ 
51       & 2023-08-26 & 06:42 & 160 &  8.4 & H & $-1.72 \pm 0.10\%$ & $ 84.7 \pm   1.7$\\ 
64       & 2023-01-14 & 06:20 & 160 &  5.5 & J & $-0.45 \pm 0.10\%$ & $ 93.0 \pm   6.5$\\ 
64       & 2023-01-14 & 06:29 & 160 &  5.5 & H & $-0.40 \pm 0.10\%$ & $ 91.9 \pm   7.4$\\
72       & 2022-07-11 & 07:02 & 160 &  6.3 & H & $-2.61 \pm 0.11\%$ & $ 90.5 \pm   1.1$\\ 
72       & 2022-07-11 & 07:07 & 160 &  6.3 & J & $-1.78 \pm 0.11\%$ & $ 89.0 \pm   1.7$\\
72       & 2022-10-15 & 03:34 & 960 & 29.7 & H & $+2.22 \pm 0.12\%$ & $  3.2 \pm   1.4$\\ 
72       & 2022-10-15 & 03:45 & 960 & 29.7 & J & $+2.07 \pm 0.11\%$ & $  0.1 \pm   1.4$\\ 
72       & 2022-12-01 & 02:29 & 960 & 27.0 & H & $+1.31 \pm 0.11\%$ & $  7.4 \pm   2.3$\\ 
72       & 2022-12-01 & 02:40 & 960 & 27.0 & J & $+1.04 \pm 0.11\%$ & $  1.5 \pm   2.9$\\ 
72       & 2023-08-26 & 11:49 & 1920 & 22.4 & J & $-0.46 \pm 0.13\%$ & $ 56.1 \pm  14.2$\\ 
72       & 2023-08-26 & 12:12 & 1920 & 22.4 & H & $+0.26 \pm 0.12\%$ & $ 11.9 \pm  11.5$\\ 
72       & 2024-01-29 & 06:52 & 960 & 13.5 & J & $-2.21 \pm 0.10\%$ & $ 89.0 \pm   1.3$\\ 
72       & 2024-01-29 & 07:15 & 960 & 13.5 & H & $-1.62 \pm 0.11\%$ & $ 91.4 \pm   1.8$\\ 
72       & 2025-04-12 & 08:20 & 160 &  8.0 & J & $-2.04 \pm 0.11\%$ & $ 86.2 \pm   1.5$\\ 
72       & 2025-04-12 & 08:29 & 160 &  8.0 & H & $-2.43 \pm 0.11\%$ & $ 88.3 \pm   1.2$\\ 
72       & 2025-05-28 & 05:31 & 140 & 15.9 & J & $-1.85 \pm 0.11\%$ & $ 87.5 \pm   1.7$\\ 
72       & 2025-05-28 & 05:38 & 160 & 15.9 & H & $-1.66 \pm 0.12\%$ & $ 87.4 \pm   1.8$\\
88       & 2023-07-26 & 09:21 & 160 & 22.1 & J & $+1.00 \pm 0.10\%$ & $  0.1 \pm   3.0$\\ 
88       & 2023-07-26 & 09:30 & 160 & 22.1 & H & $+0.85 \pm 0.10\%$ & $170.1 \pm   3.6$\\ 
88       & 2023-08-26 & 06:52 & 160 & 12.6 & J & $-1.02 \pm 0.10\%$ & $ 92.4 \pm   2.9$\\ 
88       & 2023-08-26 & 07:01 & 160 & 12.6 & H & $-0.85 \pm 0.10\%$ & $ 87.4 \pm   3.4$\\ 
88       & 2024-01-29 & 03:03 & 160 & 19.7 & H & $+0.45 \pm 0.12\%$ & $ 10.6 \pm   8.0$\\ 
88       & 2024-01-29 & 03:12 & 160 & 19.7 & J & $+0.33 \pm 0.11\%$ & $169.9 \pm  12.0$\\ 
97       & 2025-03-02 & 09:44 & 160 &  6.1 & H & $-1.67 \pm 0.11\%$ & $ 89.6 \pm   1.7$\\ 
97       & 2025-03-02 & 09:52 & 160 &  6.1 & J & $-1.60 \pm 0.11\%$ & $ 83.3 \pm   1.9$\\ 
140      & 2023-07-26 & 10:53 & 960 & 25.7 & J & $-0.14 \pm 0.11\%$ & $116.5 \pm  20.8$\\ 
140      & 2023-07-26 & 11:16 & 960 & 25.7 & H & $+0.35 \pm 0.10\%$ & $  0.8 \pm   8.7$\\ 
140      & 2023-08-26 & 08:30 & 320 & 20.6 & J & $-0.76 \pm 0.11\%$ & $ 93.7 \pm   4.2$\\ 
140      & 2023-08-26 & 08:39 & 320 & 20.6 & H & $-0.72 \pm 0.11\%$ & $ 85.9 \pm   4.5$\\ 
145      & 2021-11-08 & 07:02 & 1920 &  6.7 & J & $-1.60 \pm 0.10\%$ & $ 90.2 \pm   1.8$\\ 
145      & 2021-11-08 & 07:13 & 1920 &  6.7 & H & $-1.78 \pm 0.10\%$ & $ 92.6 \pm   1.6$\\ 
145      & 2023-01-07 & 12:29 & 960 & 23.9 & J & $+1.25 \pm 0.11\%$ & $175.2 \pm   2.4$\\ 
145      & 2023-01-07 & 12:53 & 960 & 23.9 & H & $+1.24 \pm 0.12\%$ & $178.2 \pm   2.4$\\ 
145      & 2023-05-16 & 06:43 & 120 & 14.1 & J & $-1.23 \pm 0.14\%$ & $ 84.1 \pm   3.5$\\ 
172      & 2024-01-29 & 11:17 & 960 & 15.7 & J & $-0.88 \pm 0.12\%$ & $ 93.2 \pm   3.7$\\ 
172      & 2024-01-29 & 11:40 & 960 & 15.7 & H & $-1.25 \pm 0.12\%$ & $ 83.8 \pm   2.4$\\ 
173      & 2023-08-26 & 10:48 & 160 & 26.9 & J & $+2.76 \pm 0.11\%$ & $177.2 \pm   1.2$\\ 
173      & 2023-08-26 & 10:56 & 160 & 26.9 & H & $+2.16 \pm 0.12\%$ & $  2.4 \pm   1.5$\\ 
173      & 2023-11-08 & 12:44 & 160 & 17.0 & J & $-0.68 \pm 0.11\%$ & $ 87.1 \pm   4.4$\\ 
173      & 2023-11-08 & 12:52 & 160 & 17.0 & H & $-0.49 \pm 0.12\%$ & $ 77.1 \pm   6.0$\\ 
173      & 2024-01-29 & 06:07 & 160 & 19.9 & J & $+0.22 \pm 0.11\%$ & $176.7 \pm  15.1$\\ 
173      & 2024-01-29 & 06:15 & 160 & 19.9 & H & $+0.15 \pm 0.12\%$ & $  6.6 \pm  21.8$\\ 
173      & 2025-03-02 & 10:01 & 160 &  3.3 & J & $-1.03 \pm 0.13\%$ & $ 92.3 \pm   3.6$\\ 
173      & 2025-03-02 & 10:10 & 160 &  3.3 & H & $-0.99 \pm 0.11\%$ & $ 75.1 \pm   4.2$\\
234      & 2023-07-26 & 06:51 & 160 &  6.7 & J & $-1.49 \pm 0.10\%$ & $ 89.3 \pm   2.0$\\ 
234      & 2023-07-26 & 07:00 & 160 &  6.7 & H & $-1.48 \pm 0.10\%$ & $ 93.2 \pm   2.0$\\ 
234      & 2023-08-26 & 04:08 & 160 & 18.3 & J & $-1.46 \pm 0.10\%$ & $ 90.6 \pm   2.0$\\ 
234      & 2023-08-26 & 04:17 & 160 & 18.3 & H & $-0.94 \pm 0.10\%$ & $ 91.2 \pm   3.1$\\ 
269      & 2022-05-31 & 10:52 & 160 & 17.7 & H & $-0.40 \pm 0.10\%$ & $ 74.4 \pm   7.4$\\ 
269      & 2022-05-31 & 10:57 & 160 & 17.7 & J & $-0.35 \pm 0.11\%$ & $100.3 \pm   8.5$\\ 
269      & 2022-07-11 & 08:24 & 160 &  5.7 & H & $-0.98 \pm 0.10\%$ & $ 89.1 \pm   3.1$\\ 
269      & 2022-07-11 & 08:29 & 160 &  5.7 & J & $-1.09 \pm 0.10\%$ & $ 93.7 \pm   2.8$\\
349      & 2019-03-17 & 08:05 & 1500 &  7.2 & H & $-0.36 \pm 0.10\%$ & $ 89.4 \pm   8.0$\\ 
349      & 2019-03-17 & 08:51 & 1500 &  7.2 & J & $-0.45 \pm 0.10\%$ & $ 93.7 \pm   6.6$\\ 
349      & 2021-06-26 & 11:03 & 160 & 16.6 & H & $-0.23 \pm 0.10\%$ & $ 94.5 \pm  12.7$\\ 
349      & 2021-06-26 & 11:08 & 160 & 16.6 & J & $-0.26 \pm 0.10\%$ & $ 93.9 \pm  11.4$\\ 
349      & 2021-09-04 & 06:57 & 165 & 10.2 & H & $-0.39 \pm 0.10\%$ & $ 84.4 \pm   7.5$\\ 
349      & 2021-09-04 & 07:01 & 160 & 10.2 & J & $-0.44 \pm 0.10\%$ & $ 91.8 \pm   6.7$\\ 
349      & 2021-11-08 & 02:24 & 160 & 21.4 & H & $-0.06 \pm 0.10\%$ & $109.4 \pm  46.3$\\ 
349      & 2021-11-08 & 02:29 & 160 & 21.4 & J & $+0.19 \pm 0.11\%$ & $ 26.8 \pm  15.6$\\ 
349      & 2022-10-15 & 08:42 & 160 & 17.4 & H & $-0.10 \pm 0.10\%$ & $108.6 \pm  30.4$\\ 
349      & 2023-01-07 & 06:46 & 160 & 14.2 & H & $-0.22 \pm 0.10\%$ & $ 73.5 \pm  13.0$\\ 
349      & 2023-01-07 & 06:55 & 160 & 14.2 & J & $-0.24 \pm 0.10\%$ & $ 80.0 \pm  12.1$\\ 
349      & 2023-01-14 & 05:09 & 160 & 15.9 & J & $-0.36 \pm 0.10\%$ & $ 87.1 \pm   8.2$\\ 
349      & 2023-01-14 & 05:18 & 160 & 15.9 & H & $-0.28 \pm 0.10\%$ & $101.6 \pm  10.4$\\ 
349      & 2024-01-29 & 10:28 & 160 & 10.8 & J & $-0.40 \pm 0.11\%$ & $ 83.8 \pm   7.4$\\ 
349      & 2024-01-29 & 10:38 & 160 & 10.8 & H & $-0.35 \pm 0.11\%$ & $ 93.3 \pm   8.4$\\ 
387      & 2024-01-29 & 04:26 & 960 & 18.0 & J & $-1.11 \pm 0.11\%$ & $ 87.5 \pm   3.0$\\ 
387      & 2024-01-29 & 04:49 & 840 & 18.0 & H & $-0.71 \pm 0.12\%$ & $ 86.7 \pm   5.6$\\ 
393      & 2023-05-16 & 10:08 & 140 & 24.9 & J & $+1.72 \pm 0.12\%$ & $  9.6 \pm   2.0$\\ 
393      & 2023-05-16 & 10:16 & 120 & 24.9 & H & $+1.24 \pm 0.12\%$ & $  1.6 \pm   3.0$\\ 
393      & 2023-07-26 & 06:31 & 160 & 18.3 & J & $-0.56 \pm 0.10\%$ & $ 87.6 \pm   5.5$\\ 
393      & 2023-07-26 & 06:40 & 160 & 18.3 & H & $-0.25 \pm 0.10\%$ & $ 83.3 \pm  12.0$\\ 
393      & 2023-08-26 & 03:48 & 160 & 26.0 & J & $+2.19 \pm 0.10\%$ & $  0.4 \pm   1.3$\\ 
393      & 2023-08-26 & 03:57 & 160 & 26.0 & H & $+2.20 \pm 0.11\%$ & $  1.1 \pm   1.3$\\ 
402      & 2023-07-26 & 07:30 & 160 &  3.4 & J & $-1.11 \pm 0.12\%$ & $ 89.8 \pm   3.1$\\ 
402      & 2023-07-26 & 07:39 & 160 &  3.4 & H & $-0.93 \pm 0.12\%$ & $ 94.9 \pm   3.6$\\ 
402      & 2023-08-26 & 04:43 & 960 &  9.3 & J & $-1.91 \pm 0.11\%$ & $ 88.8 \pm   1.6$\\ 
402      & 2023-08-26 & 05:06 & 960 &  9.3 & H & $-1.49 \pm 0.11\%$ & $ 90.2 \pm   2.2$\\

\hline
\end{longtable}
The degree of polarization $P_r$ has been rotated so that positive
values represent polarization perpendicular to the
Sun-Asteroid-Telescope scattering plane \textbf{at the time of the
  observations} and negative values represent polarization in this
plane. $\theta_r$ is the \textbf{measured} angle of polarization after
correcting for the observed $7.8^\circ$ systematic offset \textbf{and
  rotated so that $0^\circ$ is aligned with the vector perpendicular
  to the scattering plane}.
}
\end{center}

\clearpage

\end{document}